\documentclass[a4paper]{jpconf}

\usepackage{graphicx}

\usepackage{amsmath}

\begin{document}

\title{Charge dynamics in two-electron quantum dots}

\author{J. S\"arkk\"a and A. Harju}

\address{Helsinki Institute of Physics and Department of Applied Physics,\\ 
Aalto University, FI-02150 Espoo, Finland}

\ead{jani.sarkka(at)tkk.fi}

\begin{abstract}
We investigate charge dynamics in a two-electron double quantum dot. 
The quantum dot is manipulated by using a time-dependent external voltage that 
induces charge oscillations between the dots. We study the dependence 
of the charge dynamics on the external magnetic field and on the periodicity 
of the external potential. We find that for suitable parameter values, it is 
possible to induce both one-electron and two-electron oscillations 
between the dots.
\end{abstract}

\section{Introduction}

The search for a working realization for a quantum bit has been
going over a decade. The proposal of Loss and DiVincenzo of
using spin of a single electron trapped in a quantum dot \cite{loss98}
has been one of the most popular candidates for a qubit.
Also a setup of spins of two electrons in a quantum dot has been studied.
The quantum dot confinement potential may be realized by surface acoustic waves,
where a single or several electrons are trapped in a surface acoustic minimum.
The electrons move along the surface acoustic wave, making possible
a realization of a flying qubit \cite{shilton:531,barnes:8410,astley:156802}.
Recently Kataoka et al.~\cite{kataoka:156801} 
observed coherent dynamics of a single-electron wave function in a surface acoustic wave quantum dot. 
When the electron in a moving quantum dot enters a region with different confinement potential, the
sudden change in the potential induces excitations into the
higher energy states. This causes tunneling of electrons into an adjacent
empty channel. The charge in this channel causes a measurable current.
The rapid control of external gate voltage in two-electron double quantum dots has been demonstrated
by Petta et al.~\cite{petta05,petta10}. 

Motivated by these experiments, we study numerically charge dynamics in
two-electron quantum dots operated in a similar fashion.
Our studies show that charge dynamics may be manipulated by
choosing suitable values for parameters.

\section{Model and Method}

\begin{figure}[htb]
\begin{minipage}{11.333pc}
\includegraphics[width=11.333pc]{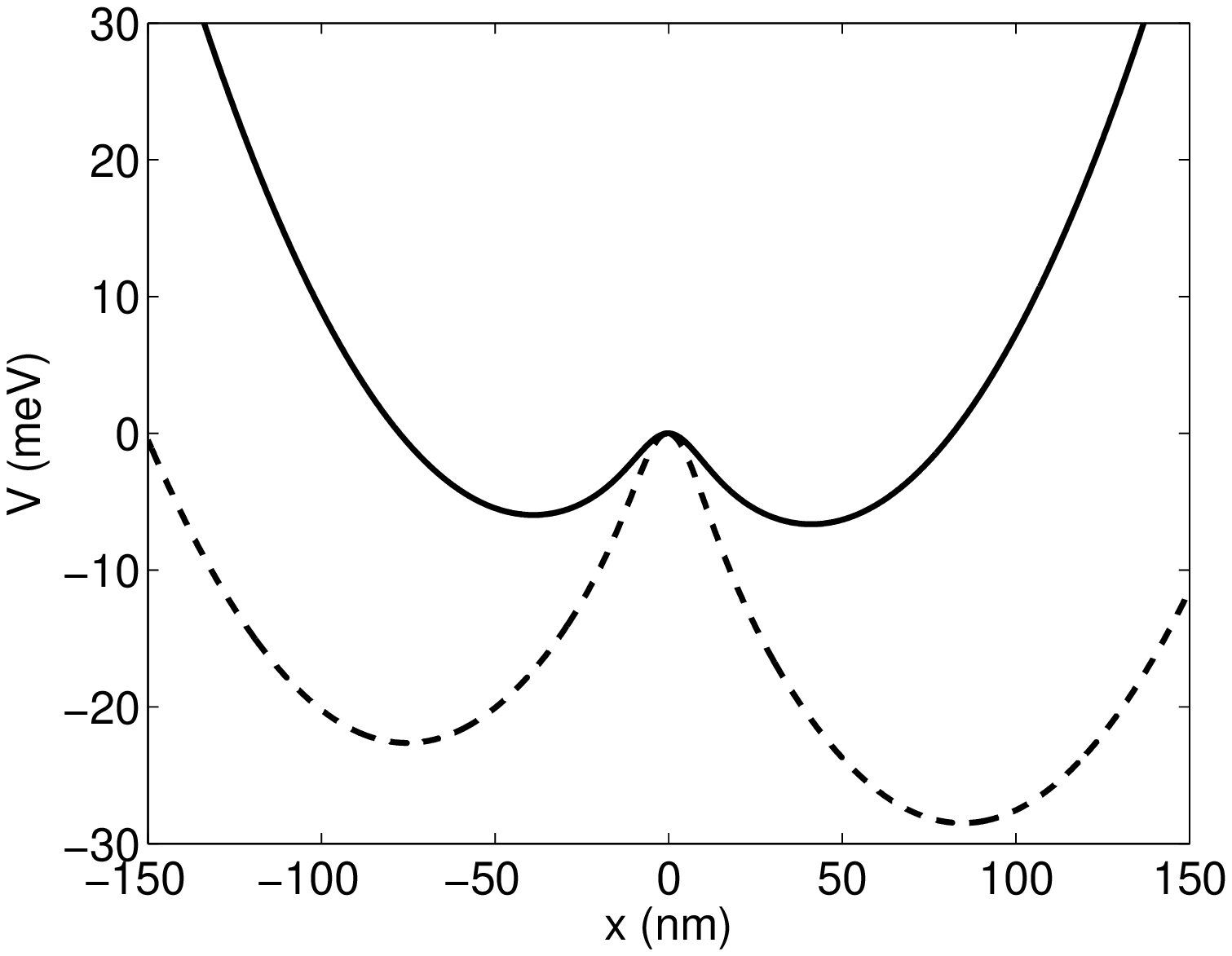}
\caption{\label{potential}
The confinement potential with $L$=80 nm, $d$=4.6 nm (dashed) and 
with $L$=40 nm, $d$=1.1 nm (solid).}
\end{minipage}\hspace{2pc}%
\begin{minipage}{11.333pc}
\includegraphics[width=11.333pc]{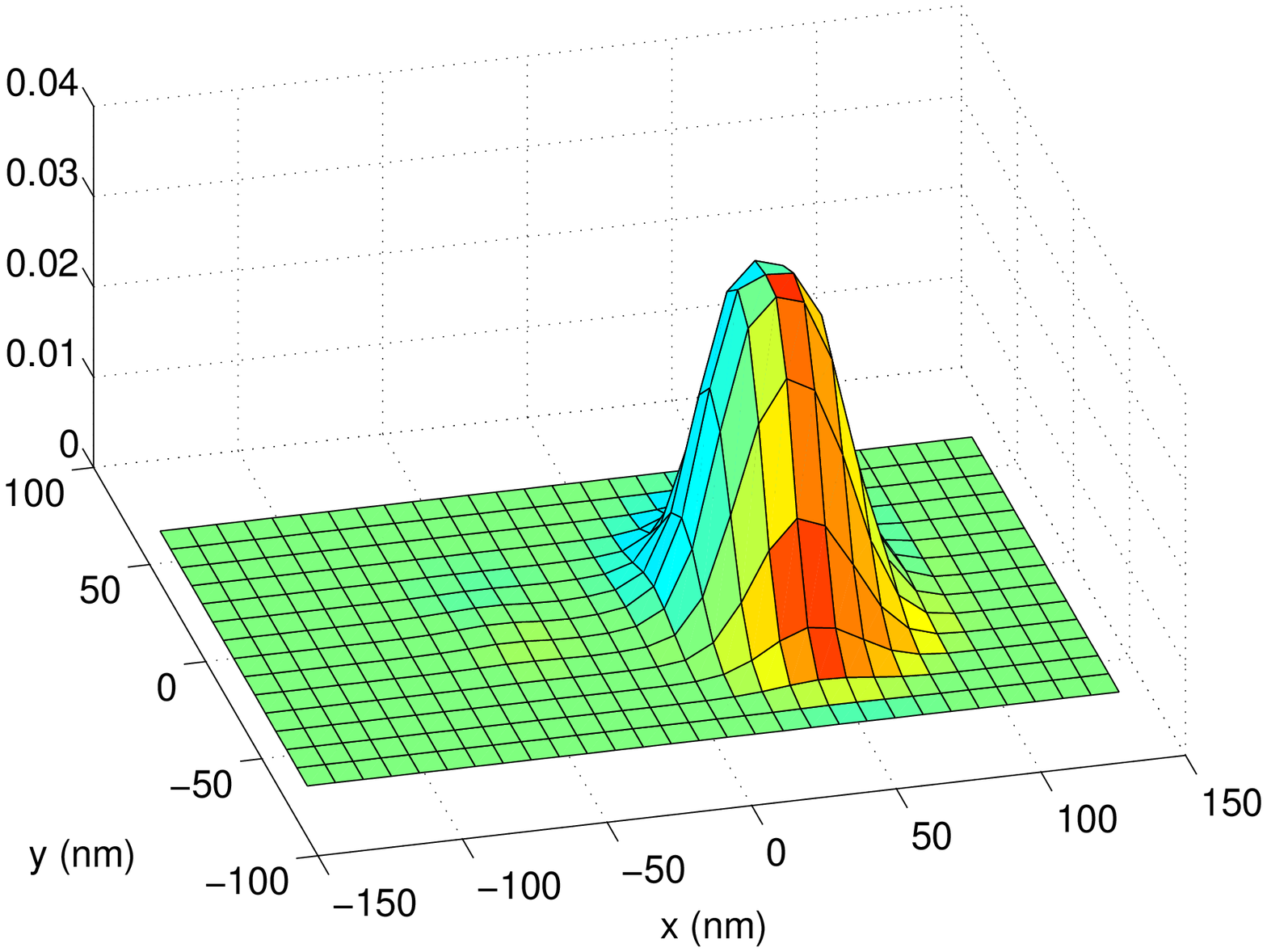}
\caption{\label{dens1}
The initial charge density when the whole charge is in
the ground state of the right dot.}
\end{minipage}\hspace{2pc}%
\begin{minipage}{11.333pc}
\includegraphics[width=11.333pc]{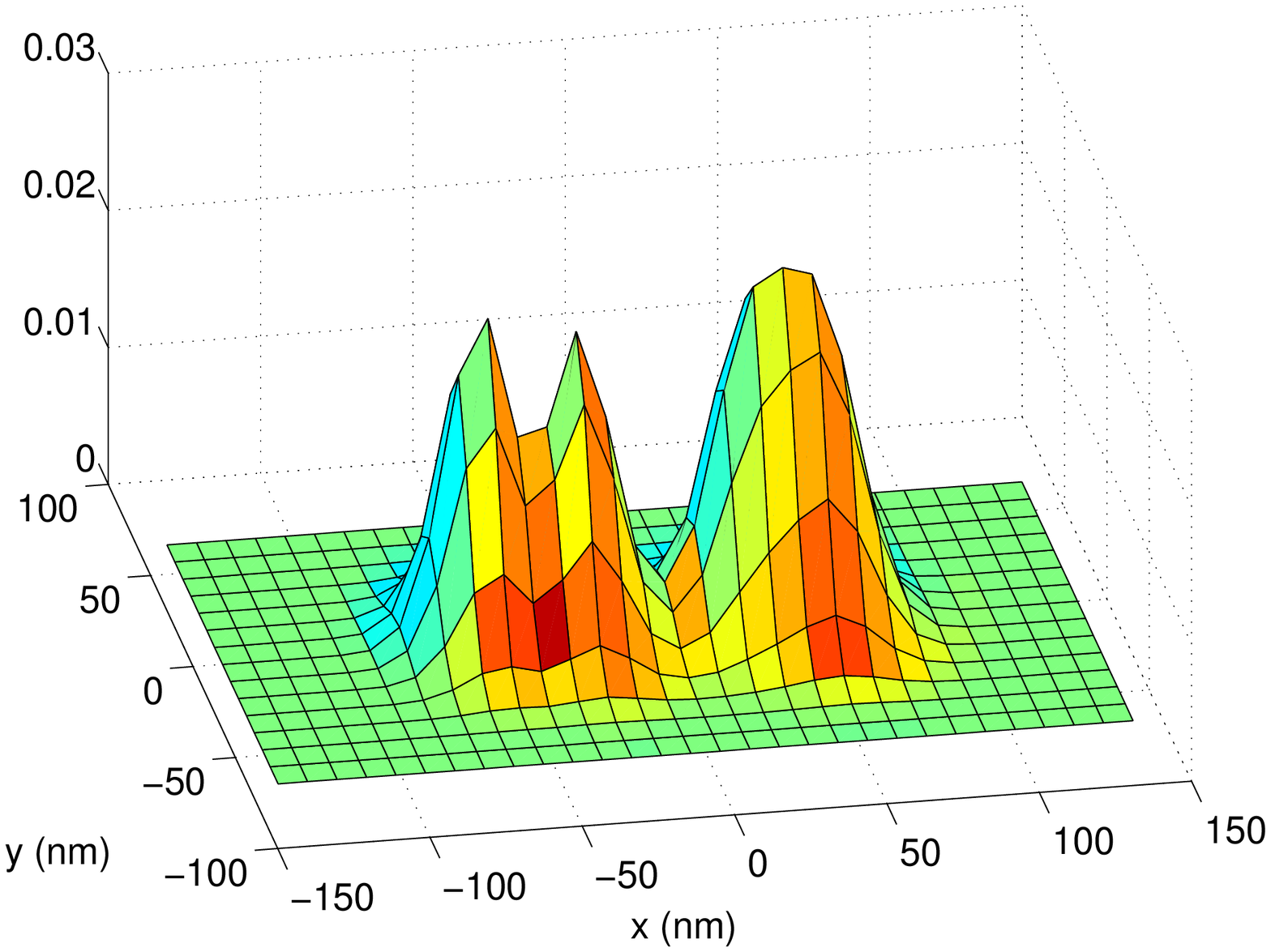}
\caption{\label{dens2}
Charge density when half of the charge is in
the excited state of the left dot.}
\end{minipage} 
\end{figure}

We model the system with two-electron Hamiltonian

\begin{equation}
\label{hamiltonian}
H=\sum_{i=1}^{2} \Bigg(\frac{\Big(-i\hbar \nabla_{i}
-\frac{e}{c}\mathbf{A}\Big)^{2}}{2m^{*}}
+V(\mathbf{r}_{i})
+g^* \mu_{B} \mathbf{B}\cdot \mathbf{s}_{i} \Bigg)
+\frac{e^{2}}{\epsilon r_{12}},
\end{equation}
where the confinement potential (illustrated in 
Fig.~\ref{potential}) is
\begin{equation}
V(\mathbf{r})=\frac{\hbar}{2}\omega^2[x^{2}-\lambda\tanh(x/\delta)x-dx]
+\frac{\hbar}{2}\omega^2 y^{2}.
\end{equation}
Here $d$ is the detuning parameter, which describes
the linear dependence of the potential on the $x$-coordinate
and has length as its unit.
The confinement strength is $\hbar \omega$= 3 meV,
$\delta$=10 nm, and the two potential minima are at points 
$x=\pm L$, $y=0$ if we set 
\begin{equation}
\lambda=L\delta/[L(1/\cosh^{2}(L/\delta))+\delta\tanh(L/\delta)].
\end{equation}

We initialize the system so that both electrons are in
the same potential minimum. This is done by setting 
the position of the potential minima to $L$=80 nm and
the initial value for the detuning parameter
$d$=4.6 nm (dashed line in Fig.~\ref{potential}), creating a ground state where both
electrons are in the right dot (see Fig.~\ref{dens1}).
Then, we lower the detuning in 5 ps to $d$=1.1 nm
and simultaneously move the positions of the potential
minima to $L$=40 nm (solid line in Fig.~\ref{potential}).
For these values, the ground state of a single electron in the right dot 
is degenerate with the first excited state of a single electron in the 
left dot (see Fig.~\ref{dens2}). Hence, charge oscillations between
the two dots are now possible.

If the detuning parameter is lowered to $d$=0 nm instead of
1.1 nm, the two-electron ground states become degenerate in both dots.
This enables the oscillation of both electrons between the dots.
We also simulate a scheme with oscillating detuning
$d$=2.8 nm+1.8 nm $\times \cos (2 \pi t/ \tau) $, where the period $\tau$
has values between 28 ps and 66 ps.

For numerical analysis of the problem, we discretize the four-dimensional
Schr\"odinger equation using finite difference method. We solve
the ground state of the system using Lanczos diagonalization. 
The time evolution $\psi(t+\Delta t)=\exp(-iH \Delta t/\hbar) \psi(t)$ 
is calculated using Krylov subspace methods 
\cite{sarkka:245315, sarkka:045323}. 

\section{Results and Discussion}

\begin{figure}[htb]
\begin{minipage}{18pc}
\includegraphics[width=18pc]{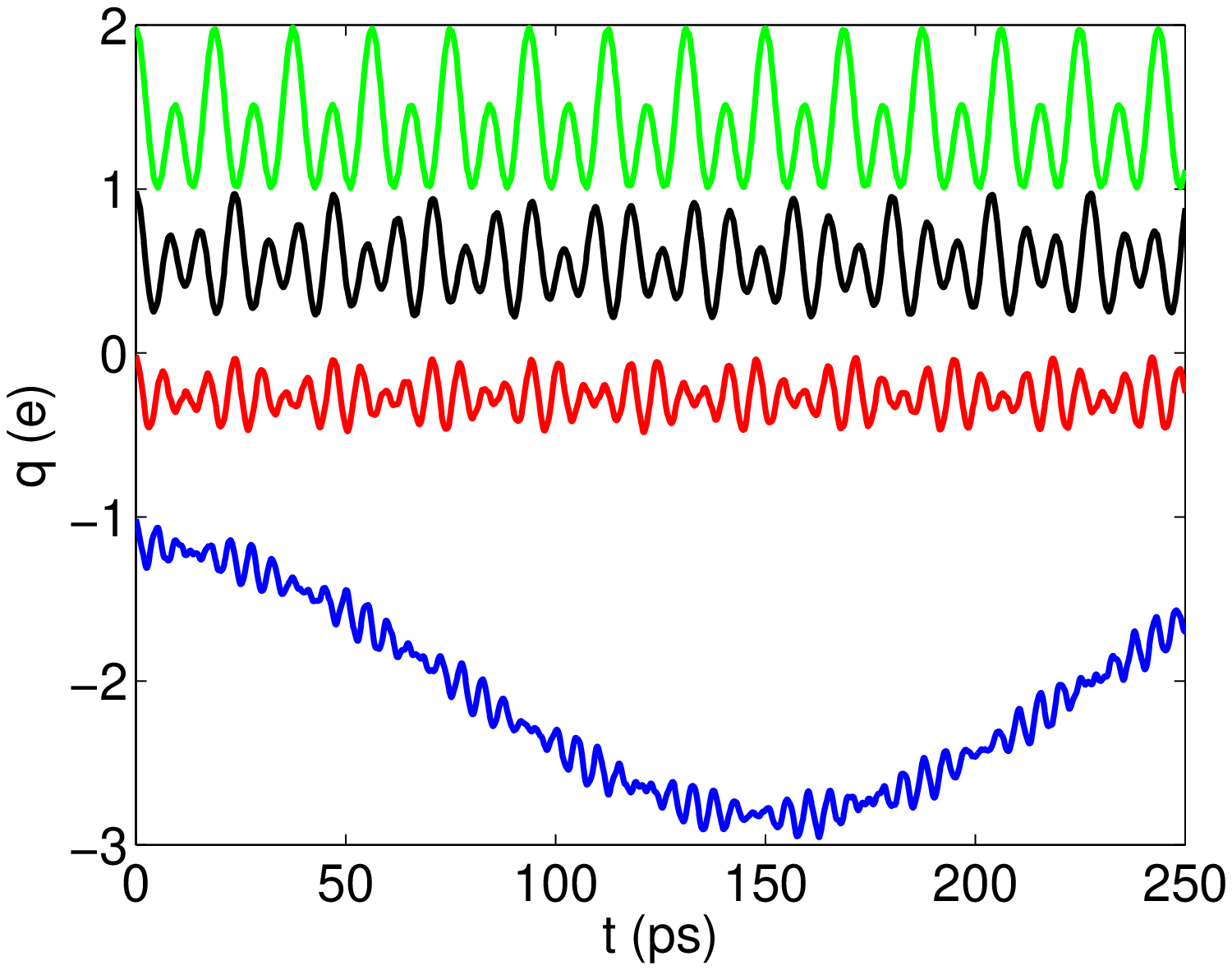}
\caption{\label{qdet}
Charge dynamics for constant detuning parameters
from down to up
\mbox{$d$=0 nm} (blue), 0.35 nm (red), 0.7 nm (black), and 1.1 nm (green).
The detuning was switched in 5 ps.
Adjacent curves below the uppermost
are offset by -1 for clarity.}
\end{minipage}\hspace{2pc}%
\begin{minipage}{18pc}
\includegraphics[width=18pc]{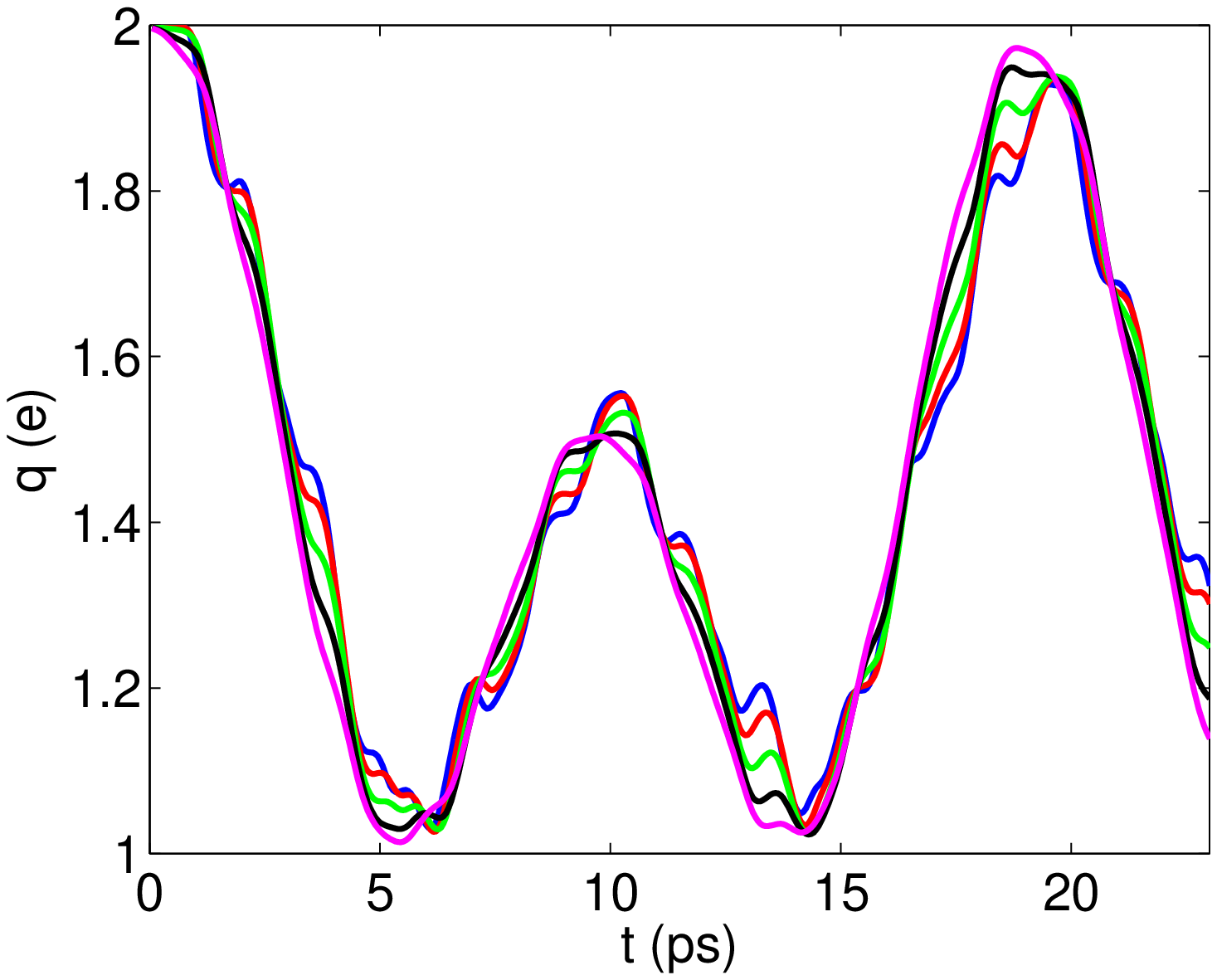}
\caption{\label{qt}
Charge dynamics for zero magnetic field and detuning
$d$=1.1 nm for different detuning switching times:
$\tau$=1 ps (blue), 2 ps (red), 3 ps (green), 4 ps (black),
and 5 ps (magenta).}
\end{minipage}
\end{figure}

We study the effect of detuning on the charge oscillations
using several detuning values, illustrated
in Fig.~\ref{qdet}. We sweep the detuning from 4.6 nm to its final value
in 5 ps, and after that keep the detuning constant in a zero magnetic field.
For zero detuning, both electrons oscillate between the dots with period 
$\tau \approx $ 300 ps, as the ground states containing two electrons in one
dot are degenerate. When the final detuning is changed to 0.35 nm,
the charge dynamics changes completely, only a fraction of the
charge goes to the other dot. This is due to the fact that
now the two-electron ground states in the dots are not degenerate.
When the final detuning is raised to 0.7 nm, more charge is moving
to the other dot. When the final detuning is 1.1 nm, 
we have a situation where the first excited state of one electron
in the first dot is degenerate with the ground state of 
two electrons in the second dot. Now one electron oscillates
between the dots with significantly faster oscillations
than in the case of degenerate ground states in the dots,
having a period $\tau$=20 ps. Using this detuning, it is
possible to induce faster charge dynamics.

The switching speed of the detuning (5 ps) is chosen so that
the behavior of the charge dynamics is sufficiently smooth.
If the switching is done in shorter time, the overall
dynamics is similar, but shorter scale
oscillations with a small amplitude emerge. This phenomenon is depicted in Fig.~\ref{qt},
where we have calculated the dynamics corresponding to the uppermost
curve in Fig.~\ref{qdet} using switching times between 1 and 5 ps.
The amplitude of the shorter scale oscillations increases slightly
with decreasing switching speed, but the period and amplitude of
the longer scale oscillations remain the same.

\begin{figure}[htb]
\begin{minipage}{18pc}
\includegraphics[width=18pc]{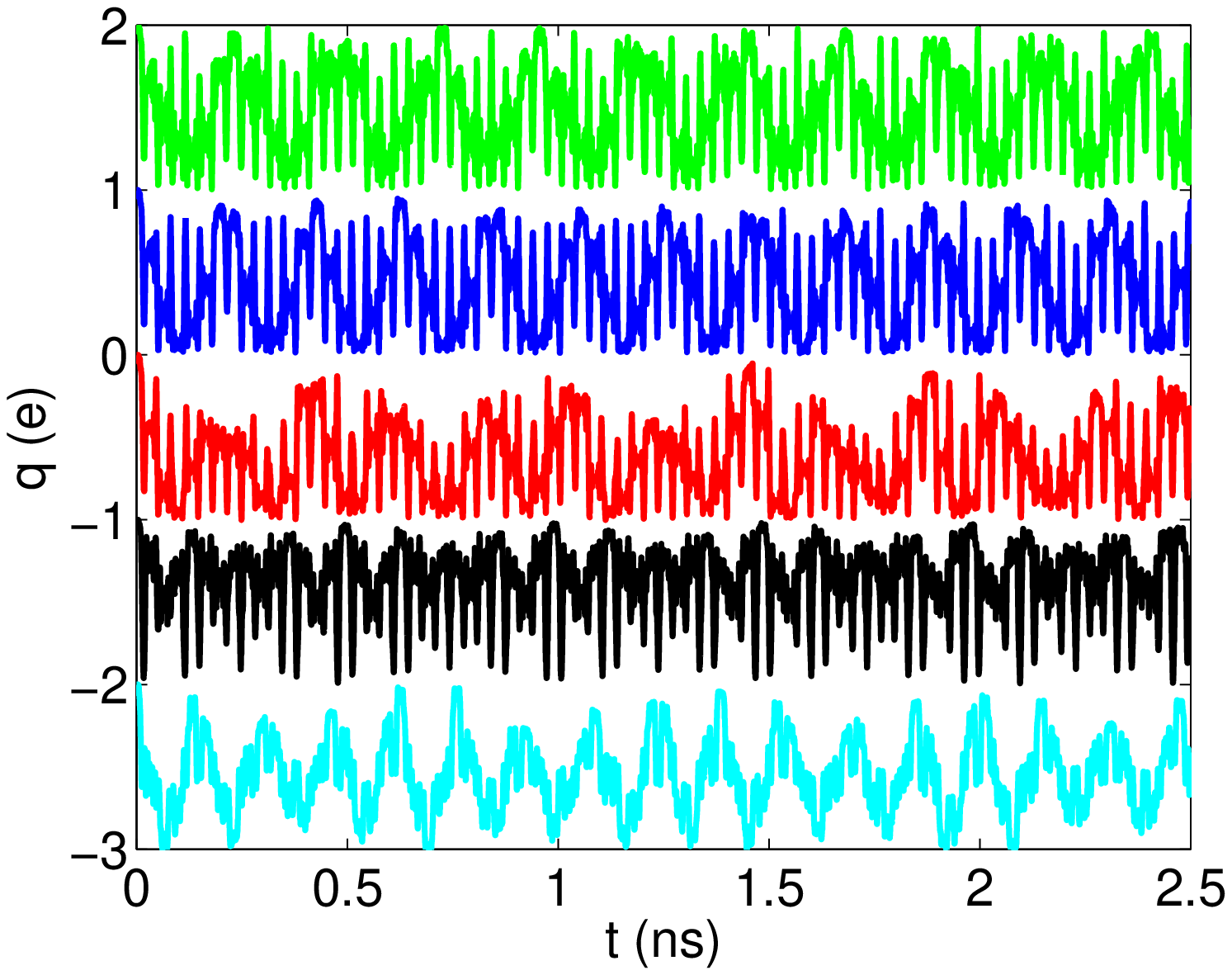}
\caption{\label{q600}
Charge dynamics for detuning oscillation period \mbox{$\tau$=33 ps}, 
detuning 1.1 nm
and different magnetic fields from up to down:
B=0 T (green), 0.03 T (blue), 0.05 T (red),
0.3 T (black), and 1 T (cyan).
Adjacent curves after the first are offset by -1 for clarity.}
\end{minipage}\hspace{2pc}%
\begin{minipage}{18pc}
\includegraphics[width=18pc]{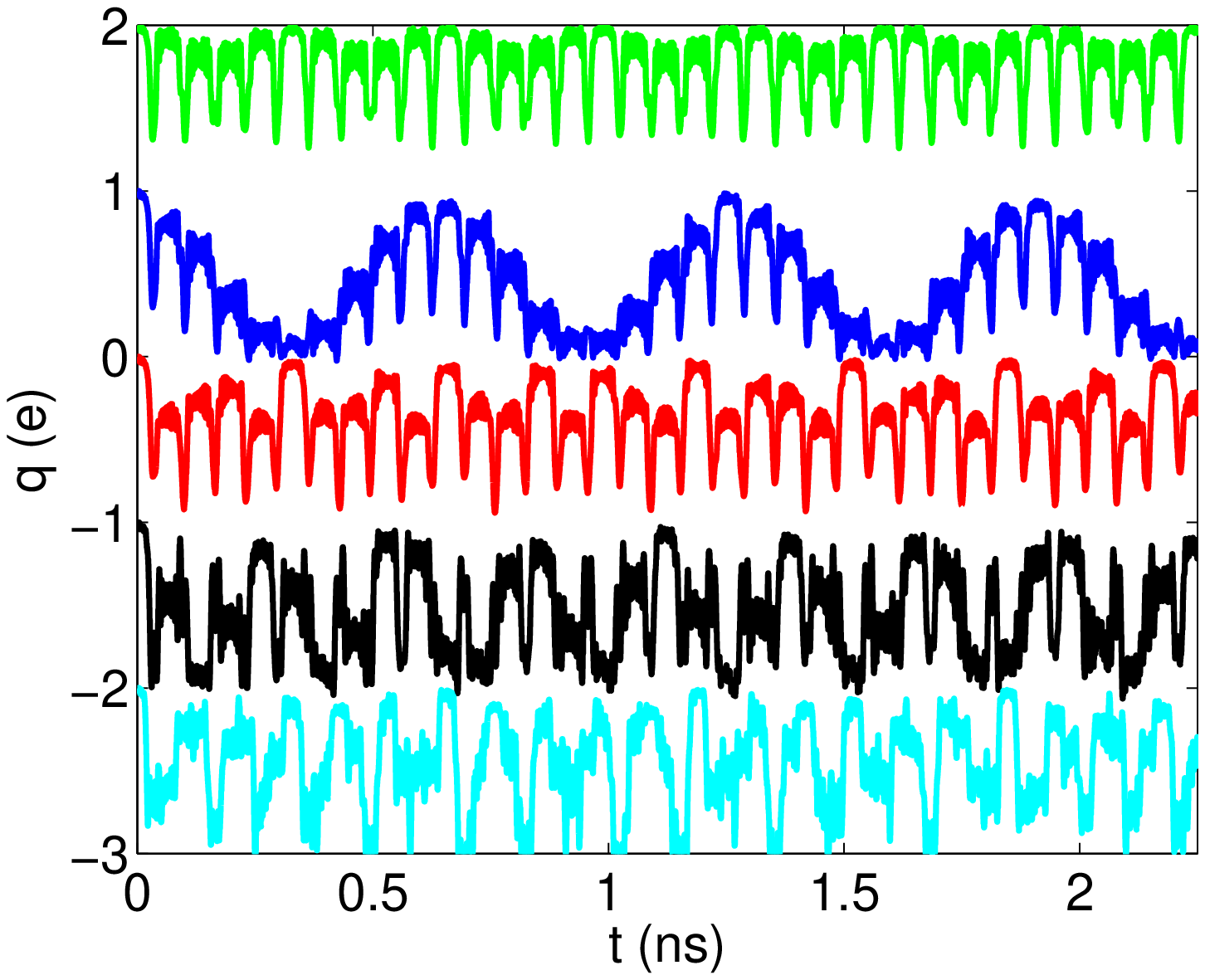}
\caption{\label{q1200}
Charge dynamics for detuning oscillation period \mbox{$\tau$=66 ps}, detuning 1.1 nm
and different magnetic fields from up to  down:
B=0 T (green), 0.03 T (blue), 0.05 T (red),
0.3 T (black), and 1 T (cyan).
Adjacent curves after the first are offset by -1 for clarity.}
\end{minipage} 
\end{figure}

Next, we apply an oscillating detuning potential and
also a non-zero magnetic field. The Zeeman term in the
Hamiltonian (Eq.~(\ref{hamiltonian})) is small compared
to the exchange energy, and the electrons
remain in the $S=0$ state.
We use two different periods for the detuning voltage, namely
$\tau$=33 ps (Fig.~\ref{q600}) and $\tau$=66 ps (Fig.~\ref{q1200}).
In Fig.~\ref{q600}, one can see that for small magnetic
fields there exists a slower oscillation in a 200 ps timescale,
and a faster oscillation having the same period as the
detuning voltage. When the magnetic field increases to 0.3 T,
the one-electron states in the dots are not degenerate anymore.
Consequently, the charge oscillations have smaller amplitude.
For a stronger magnetic field of 1 T, we find that the shorter
scale oscillations are damped, but the longer scale oscillations
have now larger amplitude.
In Fig.~\ref{q1200}, same parameters are used as in Fig.~\ref{q600},
only with the exception that the detuning oscillation period is now
two times as long. The resulting charge oscillations are
nonetheless quite different. The shorter scale oscillations have now 
a smaller amplitude, and the charge dynamics is now determined
by the longer scale oscillations. For zero magnetic field, the
charge stays mainly in the other dot. Only momentarily there
is a significant net charge in the other dot.
For magnetic field value 0.03 T, the oscillations are completely
different. Now the one-electron charge moves completely to the
other dot and back, having a period around 800 ps. 
When the magnetic field values increase, the longer scale
oscillations vanish as the system is not degenerate anymore.
The shorter-scale oscillations dominate the dynamics, which
becomes more chaotic with increasing magnetic field.
We also studied other periods for the detuning, but for values
that where not multiples of 33 ps in zero magnetic field, we did 
not observe any longer scale oscillation, and the charge dynamics was 
more chaotic. Obviously, for period 33 ps we have a kind of eigenfrequency 
for the system, although the period differs from the period 20 ps of the 
oscillations in the case of a constant detuning parameter in Fig.~\ref{qt}.

\section{Summary}
In this work, we studied charge dynamics in a two-electron
double quantum dot. We found parameters for controlling 
the detuning voltage that enable 
charge oscillations between the dots.
By calculating numerically the two-electron dynamics, we
observed two different oscillations, a slow oscillation of
two unit charges and a rapid oscillation of a single unit
charge between the dots.
The time used in switching the detuning to a constant value
did not affect the charge dynamics.
For oscillating detunings, regular and chaotic dynamics were found.

\section{Acknowledgments}
This work was supported by the Academy of Finland through its
Centers of Excellence program (2006-2011).

\section*{References}

\providecommand{\newblock}{}


\begin{thebibliography}{1}
\expandafter\ifx\csname url\endcsname\relax
  \def\url#1{{\tt #1}}\fi
\expandafter\ifx\csname urlprefix\endcsname\relax\def\urlprefix{URL }\fi
\providecommand{\eprint}[2][]{\url{#2}}

\bibitem{loss98}
Loss D and DiVincenzo D~P 1998 {\em Phys. Rev. A\/} {\bf 57} 120--126

\bibitem{shilton:531}
Shilton J~M, Talyanskii V~I, Pepper M, Ritchie D~A, Frost J~E~F, Ford C~J~B,
  Smith C~G and Jones G~A~C 1996 {\em J. Phys. Condens. Matter\/} {\bf 8}
  L531--L539

\bibitem{barnes:8410}
Barnes C~H~W, Shilton J~M and Robinson A~M 2000 {\em Phys. Rev. B\/} {\bf 62}
  8410--8419

\bibitem{astley:156802}
Astley M~R, Kataoka M, Ford C~J~B, Barnes C~H~W, Anderson D, Jones G~A~C,
  Farrer I, Ritchie D~A and Pepper M 2007 {\em Phys. Rev. Lett.\/} {\bf 99}
  156802

\bibitem{kataoka:156801}
Kataoka M, Astley M~R, Thorn A~L, Oi D~K~L, Barnes C~H~W, Ford C~J~B, Anderson
  D, Jones G~A~C, Farrer I, Ritchie D~A and Pepper M 2009 {\em Phys. Rev.
  Lett.\/} {\bf 102} 156801

\bibitem{petta05}
{Petta J R}, {Johnson A C}, {Taylor J M}, {Laird E A}, {Yacoby A}, {Lukin M D},
  {Marcus C M}, {Hanson M P} and {Gossard A C} 2005 {\em Science\/} {\bf 309}
  2180--2184

\bibitem{petta10}
{Petta J R}, {Lu H} and {Gossard A C} 2010 {\em Science\/} {\bf 327} 669--672

\bibitem{sarkka:245315}
S\"{a}rkk\"{a} J and Harju A 2008 {\em Phys. Rev. B\/} {\bf 77} 245315

\bibitem{sarkka:045323}
S\"{a}rkk\"{a} J and Harju A 2009 {\em Phys. Rev. B\/} {\bf 80} 045323

\end{thebibliography}
\end{document}